\documentstyle[12pt]{article}
\def\be{\begin{equation}}
\def\ee{\end{equation}}

\def\a{\alpha}
\def\l{\lambda}

\def\ket#1{|{#1}\rangle}  
\def\bra#1{\langle{#1}|}  
\def\norm#1#2{\langle{#1}|{#2}\rangle}   

\begin{document}
\begin{titlepage}
\rightline{IMSc/93/56}
\rightline{\em Dec 93}
\baselineskip=24pt
\begin{center}
{\large Chirality of Knots $9_{42}$ and $10_{71}$ and Chern-Simons
Theory}.

\vspace{.5cm}

{\bf P. Ramadevi, T.R. Govindarajan and R.K. Kaul} \\
The Institute of Mathematical Sciences, \\
C.I.T.Campus, Taramani, \\
Madras-600 113, INDIA.
\end{center}

\noindent {\bf Abstract}

Upto ten crossing number, there are two knots, $9_{42}$ and $10_{71}$ whose
chirality is not detected by any of the known polynomials, namely,
Jones invariants and  their two variable generalisations, HOMFLY and Kauffman
invariants. We show that the generalised knot
invariants, obtained
through $SU(2)$ Chern-Simons topological field theory, which
give the known polynomials as special cases,
are indeed sensitive to the chirality of these knots.

\vfill
\hrule
\vskip1mm
\noindent{\em email:~rama,trg,kaul@imsc.ernet.in}
\vskip1cm
\end{titlepage}

\noindent{\bf 1. Introduction}

In knot theory, one associates a polynomial with each
knot. These knot polynomials in most cases are capable of
distinguishing isotopically different knots. But there are
examples of distinct pairs of knots which have the same polynomial.
An achiral knot is that which is ambient isotopic to its mirror image.
Alexander polynomial $\Delta _K(q)$ \cite{alex} is the oldest known polynomial
but it does not
detect chirality for any knot $K$. Jones polynomial \cite {jon} $V_K(q)$
does distinguish most chiral knots. If the knot(K) has polynomial
$V_K(q)$, then the mirror image knot ($K^*$) has polynomial $V_{K^*}(q)~
=~V_K(q^{-1})$. For achiral knots we have
$V_{K}(q)~=~V_K(q^{-1})$. If $V_{K}(q)~\not =~V_K(q^{-1})$, then the knot $K$
is chiral. However the converse is not true.
There are well known examples of chiral knots which have
the polynomials symmetric under the exchange of $q \leftrightarrow
q^{-1}$.
The two variable generalisation of Jones polynomial known as HOMFLY polynomial
$P_K(l,m)$\cite{homf} is  a better invariant than Jones
polynomial since it is able to distinguish even those chiral knots
which are not distinguished from their
mirror images by Jones polynomial\cite{licko}. For these
polynomials the invariant for the mirror image is obtained
by replacing $l$ by $l^{-1}$. For example
the arborscent knot whose chirality is detected by HOMFLY
(i.e $P_K(l,m) \ne P_{K^*}(l,m)~=~P_K(l^{-1},m)$)
but not by Jones polynomial. Another two variable
generalisation of Jones invariant known as Kauffman polynomial ($F_K(\a,q)$)
\cite{onknots}is shown to detect
chirality of all those knots detected by HOMFLY\cite{kauff}.
This polynomial in fact is more powerful. For example, while the HOMFLY
invariant does not detect chirality of knot $10_{48}$ (as listed in knot
tables in Rolfsen's book\cite {rolf}), but the
Kauffman invariant does
\cite {kauff},
i.e., $F_K(\a,q) \not = F_{K}(\a ^{-1},q)$ but $P_{10_{48}}(l,m)~=
{}~P_{10_{48}}(l^{-1},m)$.

 Recently Chern-Simons field theory on arbitrary 3-manifolds
has been used to study knot invariants(\cite {wit}-\cite{gud}).
Specifically the
 expectation values of Wilson loop operators, which are the observables
 of the theory, are topological invariants.
Witten\cite{wit} has explicitly shown that Wilson loop operators
associated with skein related links
($V_+, V_-, V_0$)
in an $SU(2)$ Chern-Simons theory with doublet representation living on
the knots satisfy the same skein recursion relation as the Jones
polynomials.
Further the
 recursion relation for the two-variable generalisation (HOMFLY)
is obtained by considering
$N$-dimensional representation living on the Wilson lines in $SU(N)$
 Chern-Simons theory. Using $SO(N)$ Chern-Simons theory, Wu and
Yamagishi\cite {kengo} obtained the skein relation for Kauffman
 polynomial by placing the $N$ dimensional representation of $SO(N)$
on the
 Wilson lines.
Another approach for obtaining the link invariants involves
statistical models and Yang-Baxter equations. Using this
approach Akutsu and Wadati\cite {akut} obtained Jones polynomial and a
class of generalisation through $N$ state vertex models. In fact
Wu and Yamgishi showed that the Kauffman polynomial for the group
$SO(3)$ is same as that of Akutsu-Wadati polynomial
obtained from 3 state vertex model.

Following Witten, we have developed techniques
for obtaining new knot invariants by placing arbitrary
 representations on the Wilson lines in the
 $SU(2)$ or $SU(N)$ Chern-Simons theories\cite {ours}.
 When a 3-dimensional representation is placed on the
lines of Wilson loops in an $SU(2)$ Chern-Simons theory,
the polynomial coincides
 with Akutsu-Wadati
 polynomials derived from 3-state model. This is
so because 3 dimensional representation of $SU(2)$ is same as the
fundamental representation for $SO(3)$.

Though HOMFLY and Kauffman two-variable invariants are more powerful
than those of Jones, yet there are examples of isotopically distinct
knots which have the same polynomial. In particular,
there are two knots, namely $9_{42}$ and $10_{71}$, upto ten crossing
number whose chirality is
not detected by any of the
well known polynomials, namely, Jones, HOMFLY and Kauffman.
Using our direct method of evaluation from
$SU(2)$ Chern-Simons theory, we have explicitly derived the
knot polynomial formulae for $9_{42}$ and $10_{71}$ for
arbitrary representations
of $SU(2)$. Using macsyma package, a general algorithm has been written
to compute these
formulae.
We have verified that the fundamental representation gives Jones
polynomials and the 3 dimensional representation gives Akutsu-Wadati
/Kauffman polynomials. The 4 dimensional representation gives
polynomials which are not invariant under the transformation of the
variable $q \leftrightarrow q^{-1}$. Hence these polynomials
distinguish the
knots $9_{42}$ and $10_{71}$ from their mirror images.

 In sec.2, we give a brief account of the known polynomials for the knot
 $9_{42}$ and $10_{71}$. In sec.3, we recapitulate the necessary
ingredients of our
direct method from $SU(2)$ Chern-Simons theory\cite{ours} and do the
evaluation
for these specific knots in detail.
 In sec.4, we summarize our results.
\vskip1cm
\noindent{\bf 2. Known invariants for knots $9_{42}$ and $10_{71}$.}

As stated earlier, knots $9_{42}$ and $10_{71}$ are special knots whose
chirality is not detected by any of the known polynomials, Jones, HOMFLY
and Kauffman. We now list these polynomials for these knots.

We begin with the
knot $9_{42}$. This is a non-alternating knot with writhe $-1$. It has
signature $2$ and thus is
a chiral knot. This knot can be represented as the closure of
a four strand braid given in terms of the generators $b_1$, $b_2$, $b_3$
as a word:
$b_1^3 ~b_3~ b_2^{-1}~b_3~b_1^{-2}~b_2^{-1} $. This is drawn in fig.1a.
An equivalent representation has
been drawn in fig.1b.
Jones polynomial for this knot can be recursively obtained using the
skein recursion
relation \cite {jon}: $q^{-1}V_+~-~qV_-~=~(q^{-1/2}-~q^{1/2})V_0$.
Here $V_-,~ V_+,~ V_0$ denote the polynomials for the skein-triplet.
In the case
here, $V_-$ corresponds to $9_{42}$, $V_+$
represents the unknot obtained by changing the
encircled undercrossing of fig.1b to overcrossing
and $V_0$ obtained by changing the undercrossing
to no crossing.
The Jones polynomial for $9_{42}$ with the normalisation for unknot as
$V_U(q)~=~1$ can easily be worked out \cite{jon}:
\be
V_{9_{42}}(q) = q^{-3} - q^{-2} + q^{-1} -1 + q - q^2 +q^3 \label
{jones}
\ee
Notice that $V_{9_{42}}(q)=V_{9_{42}}(q^{-1})$ and hence this polynomial
does not detect
the chirality of $9_{42}$.

In a similar fashion, we can use the skein relation for the 2-variable
generalization of Jones invariant
referred to as HOMFLY polynomial to get\cite {licko}:
\be
P_{9_{42}}(l,m)=(-2l^{-2}-3-2l^2)+(l^{-2}+4+l^2)m^2-m^4
\ee
Clearly the polynomial is invariant under the transformation of $l\rightarrow
l^{-1}$  which relates the knot invariant to that of its mirror image.
Hence this polynomial also does not distinguish $9_{42}$ from its mirror
image.

Kauffman has obtained another
two-variable generalization of Jones invariant
through a new recursion relation for unoriented
knots or links. This polynomial for knot $9_{42}$ is \cite {onknots}:
$$
F_{9_{42}}(a,z)=(a+a^{-1})z^7 + (a^1 + a^{-1})^2z^6 -5(a+a^{-1})z^5
-5(a^1 +a^{-1})^2z^4
\hskip2cm $$
\be
\hskip1.5cm~~~~+6(a + a^{-1})z^3 +
(6a^2+12+6a^{-2})z^2-
2(a+a^{-1})z^1
-(2a^2+3+2a^{-2})z^0 \label {kau}
\ee
Kauffman polynomials for mirror reflected knots are obtained by $a
\leftrightarrow a^{-1}$. Clearly, the invariant in eqn(\ref {kau}) does not
change under this transformation and hence even this polynomial is not
powerful enough.

For the special case of $a~=-q^{-{3\over 4}}$ and $z~=q^{-{1\over 4}}+
q^{{1\over 4}}$ this polynomial reduces
to the Jones polynomial.
Also Kauffman invariants, in general reduce to Akutsu-Wadati polynomial
for
$a~=iq^2$ and $z~=-i(q-q^{-1})$. In their original calculations obtained
from 3-state vertex model, Akutsu, Deguchi and Wadati\cite {degu}
have presented these polynomials for knots with representation in terms
of closure of three strand braids. The Knot $9_{42}$, however, has a
representation in terms of closure of a four strand braid. Thus
substituting $a=iq^2$ and $z=-i(q-q^{-1})$ in eqn(\ref{kau}) should
yield us Akutsu-Wadati polynomial for this knot:
$$
F_{9_{42}}(iq^2,-i(q-q^{-1}))~= q^{-10}-q^{-9}-q^{-8}+2q^{-7}-q^{-6}-
q^{-5}+2q^{-4}
-q^{-3}+q^{-1} $$
\be
{}~\hskip3cm -1+q^{1}-q^{3}+2q^{4}-q^{5}-q^{6}+2q^{7}-q^{8}
-q^{9}+q^{10}\label {kauf}
\ee

Now let us list the
known polynomials for the other chiral knot we study here, $10_{71}$. This is
an alternating knot with writhe number zero and also signature zero.
Its knot diagram as given in Rolfsen's book\cite{rolf} is drawn in fig.2. Its
Jones polynomial is \cite{jon}:
\be
V_{10_{71}}(q)=-q^5+3q^5-6q^3+10q^2-12q+13-12q^{-1}+10q^{-2}-
6q^{-3}+3q^{-4}-q^{-5} \label {k107j}
\ee

The Kauffman two-variable polynomial for this knot can be obtained
readily from Kauffman's recursion relations. Such an exercise leads us
to:
\begin{eqnarray*}
F_{10_{71}}(a,z)&=&(a+a^{-1})z^9+(3a^2+6+3a^{-2})z^8+
(4a^3+8a+8a^{-1}+4a^{-3})z^7\\&&+
(3a^4+2a^2-2+2a^{-2}+3a^{-4})z^6+
(a^5-5a^3-15a-15a^{-1}-5a^{-3}+a^{-5})z^5\\&&-
(6a^4+12a^2+12+12a^{-2}+6a^{-4})z^4+
(-2a^{-5}+7a^{-1}+7a^{-1}-2a^{-5})z^3\\&&+
(4a^4+10a^2+12+10a^{-2}+4a^{-4})z^2+
(a^5+a^3-a-a^{-1}+a^{-3}+a^{-5})z^1
\end{eqnarray*}
\be
-(a^4+3a^2+3+3a^{-2}+a^{-4})z^0\hskip4.5cm
\ee
Again, when we substitute $a=iq^2$, $z=-i(q-q^{-1})$, we obtain the
Akutsu-Wadati invariant:
\begin{eqnarray*}
F_{10_{71}}(iq^2,-i(q-q^{-1}))&=&
q^{15}-3q^{14}+q^{13}+9q^{12}-17q^{11}+q^{10}+37q^{9}-47q^8-12q^7\\&~&+
89q^6-77q^5-42q^4+140q^3-87q^2-73q+161-73q^{-1}\\&~&-87q^{-2}+
140q^{-3}-42q^{-4}-77q^{-5}+89q^{-6}-
12q^{-7}-47q^{-8}
\end{eqnarray*}
\be
\hskip3.5cm +37q^{-9}+q^{-10}-17q^{-11}+9q^{-12}+q^{-13}-3q^{-14}+q^{-15}
\label{k107a}
\ee
Clearly, none of these invariants distinguishes $10_{71}$ from its
mirror image.

All these invariants
can also be obtained from Chern-Simons theories. Such
theories, besides these yield a whole variety of new invariants
which are powerful enough to detect the chirality of knots $9_{42}$ and
$10_{71}$.
\vskip1cm
\noindent{\bf 3. Invariants through Chern-Simons theory.}

To evaluate new Chern-Simons invariants of knots $9_{42}$
and $10_{71}$, we shall now
describe briefly the necessary aspects of the method\cite {wit, ours}.

The metric independent action of Chern-Simons theory in a 3-manifold
 is given by
\be
kS={k\over 4 \pi} \int  tr(AdA + {2\over 3}A^3)
\ee
where $A$ is matrix valued gauge connection of the gauge group G which
for our present discussion is $SU(2)$. The
gauge invariant operators of this topological field theory are given in
terms of Wilson loop(knot) operator:
$$W_R[C] = tr _R P \exp \oint _C A$$
for an oriented knot $C$ carrying representation $R$ of the group G.
The vacuum expectation value for this operator is given in terms of the
functional integral:
\be
V_R[C] = \langle W_R[C] \rangle = {\int [dA]  W_R[C]
\exp {ikS}
\over \int [dA] \exp {ikS}} \label {knot}
\ee
Being obtained from a topological field theory this quantity
depends only on the topological properties of the knot and
not on the geometric properties.

The knot invariants in eqn.(\ref {knot}) can be calculated by using a
close relationship between Chern-Simons theories on a three-manifold with
boundary and the corresponding Wess-Zumino conformal field theory on that
boundary\cite {wit}-\cite {labas}. The Chern-Simons functional integral over a
manifold is represented as a vector in the Hilbert space of the
conformal blocks associated with the boundary. For example, for a
three-ball containing two Wilson lines as shown in fig.3a, the functional
integral is a vector $\ket{\psi _0}$ in the Hilbert space ${\cal H}$
associated with boundary, four-punctured $S^2$. Functional integral over
the same ball but with opposite orientation of its boundary (fig.3b) is
represented by the state $\bra {\psi _0}$ in the dual Hilbert space $\bar
{\cal H}$. The dimension of the space is given by the number of 4-point
conformal blocks on $S^2$. For our purpose where spin ${n\over 2}$
($n+1$ dimensional
representation) live on the punctures, the dimensionality of the
Hilbert space is
$min(n,k-n)+1$. These states can be expanded in terms of a complete set of
basis.
Two convenient choices of these bases are those in which the braid
matrix for side two strands or central two strands is diagonal, $\ket
{\phi_l^{side}}$ or $\ket {\phi_l^{cent}},l~=~ 0,1,\cdots ,min(n,k-n)$. These
basis vectors
correspond to two equivalent sets of conformal blocks for four-point
correlators in $SU(2)_k$ Wess-Zumino model. These bases are related to
each other by orthogonal duality matrices as depicted below. The
duality matrices $a_{j\ell}$ are given in terms of quantum Racah coefficients
as \cite{ours}:
\setlength{\unitlength}{1cm}
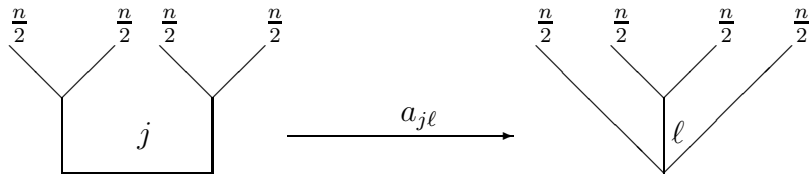
$$
\begin{picture}(11,3)
\put(1,0){\line(1,0){2}}
\put(1,0){\line(0,1){1}}
\put(1,1){\line(1,1){.7}}
\put(0.3,1.7){\line(1,-1){.7}}
\put(3,0){\line(0,1){1}}
\put(3,1){\line(1,1){.7}}
\put(2.3,1.7){\line(1,-1){.7}}
\put(4,.5){\vector(1,0){3}}
\put(9,0){\line(0,1){1}}
\put(9,1){\line(1,1){.7}}
\put(8.3,1.7){\line(1,-1){.7}}
\put(9,0){\line(1,1){1.7}}
\put(7.3,1.7){\line(1,-1){1.7}}
\put(5.5,.7){$a_{j\ell}$}
\put(.3,1.9){${n\over 2}$}
\put(1.7,1.9){${n\over 2}$}
\put(2,.4){$j$}
\put(3.7,1.9){${n\over 2}$}
\put(2.3,1.9){${n\over 2}$}
\put(9.1,.4){$\ell$}
\put(7.3,1.9){${n\over 2}$}
\put(8.3,1.9){${n\over 2}$}
\put(9.7,1.9){${n\over 2}$}
\put(10.7,1.9){${n\over 2}$}
\end{picture}
$$
where $$a_{j\ell}  =  (-)^{\ell+j-n} \sqrt{[2j+1][2\ell+1]} \left( \matrix {n/2
& n/2 &
j \cr n/2 & n/2 & \ell } \right)  $$
and the quantum Racah coefficients are given by
\begin{eqnarray*}
\pmatrix {j_{1} & j_{2} & j_{12} \cr j_{3} & j_{4} & j_{23}}&= &
\Delta(j_1,j_2,j_{12}) \Delta(j_3,j_4,j_{12}) \Delta(j_1,j_4,j_{23})
\Delta(j_3,j_2,j_{23})  \\
&&\sum_{m\geq0}{(-)^{m} [m+1]!}
\Bigl\{ {[m-j_1-j_2-j_{12}]}!\Bigr. \\
&&{[m-j_3-j_4-j_{12}]}!
{[m-j_1-j_4-j_{23}]}!\\
&&{[m-j_3-j_2-j_{23}]}!
{[j_1+j_2+j_3+j_4-m]}! \\
&&
\Bigl.{[j_1+j_3+j_{12}+j_{23}-m]}!{[j_2+j_4+j_{12}+j_{23}-m]}!\Bigr\}^{-1}
\end{eqnarray*}
and
$$ \Delta(a,b,c) = \sqrt{{{[-a+b+c]![a-b+c]![a+b-c]!}\over{[a+b+c+1]!}}}
$$
Here $[a]!= [a][a-1][a-2]...[2][1]$. The $SU(2)$ spins are related as
$\overrightarrow{j}_1 + \overrightarrow{j}_2 + \overrightarrow{j}_3 =
\overrightarrow{j}_4 , \overrightarrow{j}_1 + \overrightarrow{j}_2 =
\overrightarrow{j}_{12} , \overrightarrow{j}_2 + \overrightarrow{j}_3 =
\overrightarrow{j}_{23}$
subject to the fusion rules of $SU(2)_k$ conformal field theory.
Here the number in square bracket is the q-number
defined as $[n]={q^{n/2}-q^{-n/2}\over q^{1/2}-q^{-1/2}}$. The parameter
$q$ is related to the coupling constant $k$ of the Chern-Simons theory as
: $q=\exp ({2\pi i\over k+2})$.

The state $\ket {\psi _0}$ and its dual $\bra {\psi _0}$ corresponding
to the
fig.3a and fig.3b respectively can be written in the eigen basis discussed
above as
follows:
\begin{eqnarray}
\ket {\psi _0} = \sum_{l=0} \sqrt {[2l+1]}\ket {\phi _l^{side}}=
[n+1] \ket {\phi
_0^{cent}} \\
\bra {\psi _0} = \sum_{l=0} \sqrt{[2l+1]}\bra {\phi _l^{side}}=
[n+1] \bra {\phi
_0^{cent}}
\end{eqnarray}
 Now, glueing these two figures along
their oppositely oriented boundaries gives two disjoint unknots and
the link invariant given in eqn.(\ref {knot})
is represented by the inner product $\norm {\psi _0}{\psi _0}$. This
gives the unknot polynomial for the Wilson lines carrying
spin ${n\over 2}$ representation to be: $V_n[U]=[n+1]$.

Another building block for our calculation is the functional integral
over an $S^3$ from which two three-balls have been carved out. The two
boundaries so formed are connected by four Wilson lines as shown in
fig.4. The Chern-Simons functional integral over this three-manifold
operates like an identity and is given in terms of the basis vectors
$\ket {\phi _i^{(1)}}$ and $\ket {\phi _i^{(2)}}$ refering to the two
boundaries as\cite{ours}:
\be
\nu _2 = \sum _i \ket {\phi _i^{(1)}} \ket {\phi_i^{(2)}} \label {two}
\ee
Here the summation runs from 0 to minimum of $n$ and $k-n$ for spin ${n\over
2}$
representation living on the Wilson lines. In eqn.(\ref {two}) the basis
vectors
$\ket {\phi _i^{(1)}}$ and $\ket {\phi _i^{(2)}}$  can both refer to
either the side strands or the central strands.

Yet another useful functional integral is over an $S^3$ with three
three-balls removed from it. The consequent three boundaries (each an
$S^2$) are connected by Wilson lines as shown in fig.5. The Chern-Simons
functional integral over this manifold is given by\cite {ours}:
\begin{eqnarray}
\nu _3=\sum _{i,j,l,m=0}{1\over \sqrt {[2m+1]}}a_{im}a_{jm}a_{lm}
\ket {\phi_i^{(1)cent}}\ket
{\phi_j^{(2)cent}}\ket {\phi_l^{(3)cent}} \label {three}\\
{}~~~~=
\sum _{m=0}{1\over \sqrt {[2m+1]}}
\ket {\phi_m^{(1)side}}\ket
{\phi_m^{(2)side}}\ket {\phi_m^{(3)side}}\hskip2.5cm \label {bdy}
\end{eqnarray}
where superscripts $1,2,3$ refers to the conformal block bases on the
three $S^2$ boundaries $1,2,3$.

In the functional integral corresponding to fig.3, fig.4 and fig.5 we can
introduce any number of braids in various strands through the braiding
matrix. The braiding matrix that introduces half-twist  in the side two
strands is diagonal in the basis $\ket {\phi_l^{side}}$. On the other
hand braid matrix that twists the central two strands is diagonal in
the basis $\ket {\phi _l^{cent}}$. The eigenvalues for these braiding
matrices depend on the relative orientation  of the strands they twist.
These eigenvalues are obtained from conformal field theory and are given
for right handed half-twists in parallely and antiparallely oriented strands
respectively by \cite
{ours}:
$$\l _l^{(+)} = (-1)^{n-l}q^{(n(n+2)-l(l+1))/2}$$
$$\l _l ^{(-)}= (-1)^l q^{l(l+1)/2},~~ l~=~0,1,\cdots,min(n,k-n)$$

The properties of Chern-Simons functional integrals listed above
can now be directly
used to compute the invariants for knot $9_{42}$ and
$10_{71}$.
We begin with knot $9_{42}$. Split the
knot as represented in fig.1b at marked points 1 to 4 by vertical planes.
This breaks the manifold into five pieces as shown in fig.6a-6e. The
functional integral over each one of them can now be readily computed.
For example, the functional integral over the manifold with one boundary
in fig.6a, can be obtained by half-twisting the side two strands of fig.3a
three times
to yield:
\be
\nu _1(P_1)=\sum _{l_1=0}\sqrt
{[2l+1]}(-1)^{3(n-l_1)}q^{-3/2[n(n+2)-l_1(l_1+1)]}\ket {\phi _{l_1}^{(1)}}
\ee
Similarly, the functional integral over the manifold in fig.6e, again
with one boundary is:
\be
\nu _1(P_4)=\sum _{l_5=0}(-1)^{n-l_5}q^{-1/2[n(n+2)-l_5(l_5+1)]}\ket
{\phi _{l_5}^{(1)}}
\ee
The rest of the three manifolds have two boundaries. These can be
evaluated from fig.4 and fig.5 and using
the duality relationship between the two sets
of bases
\cite {ours}.
The state corresponding to fig.6b is
\be
\nu _2(P_1;P_2)=\sum _{i_1,j_1,l_2,r=0} {a_{l_1r}a_{j_1r}a_{l_2r}\sqrt
{[2l_2+1]}\over
\sqrt {[2r+1]}} q^{n(n+2)-l_2(l_2+1)}\ket {\phi _{i_1}^{(1)}}\ket {\phi
_{j_1}^{(2)}}
\ee
The functional integral corresponding to fig.6c is
\be
\nu _2(P_2;P_3)=\sum _{l_3=0}q^{l_3(l_3+1)}\ket {\phi _{l_3}^{(1)}}
\ket {\phi
_{l_3}^{(2)}}
\ee
Similarly the fig.6d corresponds to
\be
\nu _2(P_3;P_4)= \sum
_{i_2,j_2,l_4=0}(-1)^{l_4}q^{-l_4(l_4+1)/2}a_{l_4i_2}a_{l_4j_2}\ket
{\phi _{i_2}^{(1)}}\ket {\phi _{j_2}^{(2)}}
\ee

Glueing these five pieces (fig.6a-6e) along the appropriate oppositely
oriented boundaries gets us back to knot $9_{42}$ in $S^3$.
The
final result is :
\newpage
$$
V_n[9_{42}] = (-1)^n q^{{-3/2}n(n+2)}\sum _{r,l_1,l_2,j_1,j_2=0}\sqrt
{[2l_1+1]} \sqrt
{[2l_2+1]}\sqrt
{[2j_2+1]}\hskip1cm
$$
\be \hskip1cm a_{l_1r}a_{l_2r}a_{j_1r}a_{j_1j_2}(-1)^{l_1}q^{3/2l_1(l_1+1)}
q^{3/2j_1(j_1+1)}q^{-l_2(l_2+1)}q^{j_2(j_2+1)}
\label {result}
\ee
for spin ${n\over 2}$ representation living on the Wilson lines. In
obtaining the  final result (\ref {result}), the following identity
involving the $q$-Racah coefficient has been used\cite {ours}:
\be
\sum _{l_4} (-1)^{n-l_4}q^{(n(n+2)-l_4(l_4+1))/2} a_{j_1l_4}a_{j_2l_4}=
(-1)^{j_1+j_2}q^{(j_1(j_1+1)+j_2(j_2+1))/2}a_{j_1j_2} \label {idy}
\ee

The invariant (\ref {result}) can be evaluated explicitly for definite
values of spin ${n\over 2}$. This has been done on computer by using
macsyma package. The results are:

a) For $n=1$, we get
$$V_1[9_{42}]=q^{7/2}+q^{-7/2}$$
This is same as eq.(\ref {jones}) when divided by the unknot polynomial
$V_1[U]=q^{{1\over 2}}+q^{-{1\over 2}}$.

b) For $n=2$, we get
$$
V_2[9_{42}]=q^{11}-q^9 + q^3 +1 + q^{-3}-q^{-9}+q^{-11}
$$
This is same as Akutsu-Wadati/Kaufmann polynomial (\ref
{kauf}) when divided by the unknot polynomial $V_2[U]=q^{-1}+1+q$.

c) For $n=3$, the polynomial is
$$
V_3[9_{42}]=q^{45/2}-q^{41/2}-q^{39/2}+q^{35/2}+q^{23/2}+q^{21/2}-q^{19/2}-
q^{17/2}+q^{13/2}-q^{9/2}
$$
$$
{}~~~~~~+q^{5/2}+q^{3/2}+
q^{-3/2}+
q^{-5/2}
-q^{-13/2}-q^{-15/2}+q^{-21/2}+2q^{-23/2}
$$
\be
{}~~~~~~-q^{-27/2}
+2q^{-31/2}
-3q^{-35/2}-q^{-37/2}+q^{-39/2}+q^{-41/2}\hskip2cm
 \label{final}
\ee

Let us now study the knot $10_{71}$. The Chern-Simons functional
integral for this knot can be obtained from the functional integrals over
four 3-manifolds $I$, $II$, $III$ and $IV$ as shown in fig.7. The three
3-manifolds $I$, $II$ and $III$ are three-balls with one boundary each
($S^2$). The manifold $IV$ has three boundaries, each an $S^2$. Glueing these
pieces
together along their appropriate boundaries as shown in fig.7 yields knot
$10_{71}$ in $S^3$. The rules of obtaining Chern-Simons functional
integrals stated above can now readily be applied to obtain these
functional integrals. The final answer for the knot invariant is:
$$
V_n[10_{71}]=(-1)^n q^{{n(n+2)\over 2}}\sum _{i,r,s,u,m=0}
\sqrt {{[2r+1][2s+1][2u+1]\over
[2m+1]}}a_{im}a_{ms}a_{rm}a_{iu}~~~~~~
$$
\be
{}~~~~~~~~~~~~~~~~~~~~~~~~~~~~~~~~(-1)^s~q^{-i(i+1)}q^{m(m+1)}q^{-r(r+1)}q^{u(u+1)}
q^{{3\over 2}s(s+1)}
\ee
Here the identity given in eqn.(\ref {idy}) has again been used.

This invariant has been evaluated for specific values of the spin
${n\over 2}$ by macsyma package. For $n=1$ and $n=2$, it reduces to the
Jones and Akutsu-Wadati/Kauffman polynomials (eqn.\ref{k107j} and \ref {k107a})
upto the normalisation factor $V_1[U]=q^{{1\over 2}}+q^{-{1\over 2}}$ and
$V_2[U]=q+1+q^{-1}$ respectively. For n=3, we have the polynomial:
\begin{eqnarray*}
V_3[10_{71}]&=&-~q^{63/2}+2q^{61/2}+q^{59/2}-3q^{57/2}-4q^{55/2}+
7q^{53/2}+10q^{51/2}-15q^{49/2}\\&~&-21q^{47/2}+
22q^{45/2}+44q^{43/2}-25q^{41/2}-79q^{39/2}+17q^{37/2}+119q^{35/2}
\\&~&+8q^{33/2}-
150q^{31/2}-54q^{29/2}+172q^{27/2}+99q^{25/2}-166q^{23/2}-
144q^{21/2}\\&~&+137q^{19/2}+
174q^{17/2}-95q^{15/2}-
180q^{13/2}+46q^{11/2}+167q^{9/2}+
7q^{7/2}\\&~&-138q^{5/2}-
56q^{3/2}+101q^{1/2}+101q^{-1/2}-
56q^{-3/2}-138q^{-5/2}\\&~&+
7q^{-7/2}+168q^{-9/2}+
46q^{-11/2}-
182q^{-13/2}-96q^{-15/2}+175q^{-17/2}
\\&~&+138q^{-19/2}-144q^{-21/2}
-
165q^{-23/2}+99q^{-25/2}+171q^{-27/2}-
54q^{-29/2}\\&~&-148q^{-31/2}
+8q^{-33/2}+
115q^{-35/2}+16q^{-37/2}-77q^{-39/2}-23q^{-41/2}\\&~&+
44q^{-43/2}
+21q^{-45/2}-
21q^{-47/2}-
15q^{-49/2}+10q^{-51/2}+7q^{-53/2}
\end{eqnarray*}
\be
\qquad-4q^{-55/2}-
3q^{-57/2}+
q^{-59/2}+2q^{-61/2}-q^{-63/2}~\hskip2cm\label{final1}
\ee

Clearly, unlike Jones, HOMFLY and Kauffman/Akutsu-Wadati polynomials for knot
$9_{42}$ and $10_{71}$, the spin ${3\over 2}$ polynomials for these knots
(eqns. \ref {final} and \ref {final1} respectively) do indeed change under
chirality
tranformation  $q \leftrightarrow
q^{-1}$.
\newpage
\noindent{\bf 4. Concluding Remarks}

It appears that spin ${1\over 2}$, 1, ${3\over 2}$,$\cdots$ polynomials
are progressively more powerful.
Upto ten crossing, there are six chiral knots ($9_{42}$, $10_{48}$,
$10_{71}$,$10_{91}$, $10_{104}$ and $10 _{125}$) which are not
distinguished from their mirror images by Jones or its two-variable
generalisation, HOMFLY polynomial. Kauffman /Akutsu-Wadati polynomial is
more powerful. It does detect chirality of the knots $10_{48}$,
$10_{91}$, $10_{104}$ and $10_{125}$, but not of knots $9_{42}$ and
$10_{71}$.
We have demonstrated that the new polynomials obtained from $SU(2)$
Chern-Simons theory corresponding to spin ${3\over 2}$ living on the
knot is powerful enough to distinguish even these knots from their
mirror images.

\noindent {\bf Acknowledgements}

We would like to thank Ravi Kulkarni for bringing the chiral knot
$9_{42}$
to our notice.

\newpage
\flushleft{\bf Figure Captions:}\\
\begin{itemize}
\item[Fig.1] Knot $9_{42}$ as (a) the closure of a four-strand braid and
(b) an equivalent representation.
\item[Fig.2] Knot $10_{71}$
\item[Fig.3] Diagrammatic representation of (a) state
 $\ket {\psi _0}$ and (b) its dual $\bra {\psi_0}$ in terms of
three-balls.
\item[Fig.4] Diagrammatic representation of the functional integral $\nu_2$
for a manifold with two boundaries.
\item[Fig.5] Diagrammatic representation of the functional integral
$\nu_3$ for a manifold with three boundaries.
\item[Fig.6] Knot $9_{42}$ obtained by glueing five building blocks,
(a)--(e), with suitable entanglements.
\item[Fig.7] Knot $10_{71}$ obtained by glueing four building blocks,
(I)--(IV), with suitable braidings.
\end{itemize}

\end{document}